\documentclass[twocolumn,pra,floatfix,citeautoscript,nofootinbib,superscriptaddress]{revtex4}
\usepackage{amsbsy}
\usepackage{latexsym,epsfig,graphicx}
\usepackage{dcolumn}
\usepackage{graphicx}
\usepackage{subfigure}
\usepackage{comment}
\usepackage{color}
\usepackage{bm}
\usepackage{mathrsfs}
\usepackage{amsfonts}
\usepackage{amsmath}
\usepackage{color}
\usepackage{amssymb}
\usepackage{xspace}
\usepackage{epstopdf}
\usepackage{tabularx}
\usepackage{longtable}
\usepackage[colorlinks=true, letterpaper=true, pdfstartview=FitV, linkcolor=blue, citecolor=blue, urlcolor=blue]{hyperref}
\usepackage[normalem]{ulem}

\pdfoutput=1
\input{tcilatex}
\begin{document}

\title{Majorana corner pairs in a two-dimensional $s$-wave cold atomic
superfluid}
\author{Ya-Jie Wu}
\affiliation{Department of Physics, The University of Texas at Dallas, Richardson, Texas
75080-3021, USA}
\affiliation{School of Science, Xi'an Technological University, Xi'an 710032, China}
\author{Xi-Wang Luo}
\affiliation{Department of Physics, The University of Texas at Dallas, Richardson, Texas
75080-3021, USA}
\author{Junpeng Hou}
\affiliation{Department of Physics, The University of Texas at Dallas, Richardson, Texas
75080-3021, USA}
\author{Chuanwei Zhang}
\thanks{chuanwei.zhang@utdallas.edu}
\affiliation{Department of Physics, The University of Texas at Dallas, Richardson, Texas
75080-3021, USA}

\begin{abstract}
We propose a method to prepare Majorana pairs at the corners of imprinted
defects on a two-dimensional cold atom optical lattice with $s$-wave
superfluid pairing. Different from previous proposals that manipulate the
effective Dirac masses, our scheme relies on the sign flip of the spin-orbit
coupling at the corners, which can be tuned in experiments by adjusting the
angle of incident Raman lasers. The Majorana corner pairs are found to be
located at the interface between two regimes with opposite spin orbit
coupling strengths in an anticlockwise direction and are robust against
certain symmetry-persevered perturbations. Our work provides a new way for
implementing and manipulating Majorana pairs with existing cold-atom
techniques.
\end{abstract}

\maketitle

\section{Introduction}

Majorana zero modes (MZMs) have attracted great attention in past decades
owing to their non-Abelian exchange statistics and potential applications as
topologically protected qubits \cite{Kitaev2003,Nayak2008}. They also
exhibit significant physics in a range of disciplines such as nuclear and
particle physics \cite{Elliott2015}. Strenuous efforts to search for MZMs
are underway in both theories and experiments. In recent years, a variety of
schemes to realize Majorana excitations have been proposed \cite%
{Qix2011,Hansan2010,zhang2008,Sato2009,Jiang2011,Alicea2011,Alicea2012,Read2000,Phong2017,Oppen2010,Lutchyn2010,Jiang2016,Sau2010,Tewari2012,Teemu013}
by utilizing $p$-wave superconductors (SCs) or superfluid (SFs) \cite%
{Read2000,Phong2017}, or SCs and SFs with effective $p$-wave paring via
spin-orbit coupling (SOC) and $s$-wave pairing \cite%
{Hansan2010,Qix2011,zhang2008,Sato2009,Jiang2011}. Remarkable experimental
progresses have been made in condensed-matter systems \cite%
{Mourik2012,Finck2013,Nadj2014,Xuj2015,Wang2016,Heq2017}. The experimental
realization of SOC in ultracold atomic gases offer another clean platform to
explore Majorana physics \cite%
{Lin2011,Zhangj2012,Wang2012,Williams2013,Huang2016,Meng2016,Wuz2016}. In
these platforms, the interplay among SOC, Zeeman fields and $s$-wave
interactions could produce non-Abelian topological superfluids (TSFs) that
host Majorana excitations. There have been several tantalizing proposals for
realizing and tuning Majorana excitations, for example, by creating
topological defects (such as SF vortices or lattice dislocations) or defect
chain \cite{Lang2014,Deng2015}.

The emergence of Majorana excitations can be intuitively understood by the
low-energy theory. A pair of MZMs exist at the kinks where the pairing
potential or SOC changes the sign (which corresponds to the sign change of
Dirac mass or velocity in Jackiw--Rebbi model). In solid-state materials,
the SOC kinks are difficult to tune, while the kinks of pairing potentials
can be realized through Josephson junctions in superconducting nanowires, as
well as the corners and hinges in recently proposed higher-order topological
SCs (TSCs) \cite{J.
Langbehn2017,Benalcazar2017,Song2017,Fang2017,Schindler2018,Khalaf2017,ZhuX2018,Khalaf2018,Yan2018,Wang2018,wangyu2018,Liu2018,Hsu2018}%
. In particular, for two-dimensional (\textrm{2D}) second-order TSCs, the
bulk topology of the \textrm{2D} system offers \textrm{1D} edge modes, which
have different topologies for adjacent edges due to the change of the
pairing sign, leading to zero-energy Majorana Kramers pairs or Majorana
modes at corners \cite{Yan2018,Wang2018,wangyu2018,Liu2018}. On the other
hand, in cold atomic system, the kinks of pairing potentials may be obtained
by soliton excitations \cite{Xu2014,LiuX2015}. However, these cold atomic
systems suffer from dynamical instability in the presence of perturbations.
Thus, proposals for realizing robust MZMs in atomic systems, through other
manners like SOC kinks, are highly in demand.

In this paper, we propose feasible schemes to realize SOC kinks using
trapped ultracold fermionic atoms on a \textrm{2D} optical lattice, and show
that our system supports Majorana pairs in a vortex-free configuration. The
main results are listed below:

(i) Effective \textrm{1D} modes on a rectangular geometry would emerge in
the \textrm{2D} system through engineering on-site potential of the
rectangle. In the presence of \textrm{1D} equal Rashba-Dreeshauls (ERD) SOC,
Majorana corner pair emerges at the corner of the rectangle with a proper $s$%
-wave pairing.

(ii) Each edge of the rectangular defect is characterized by a \textrm{1D}
topological SF in the \textrm{BDI} class, and Majorana pair exists at the
interface of two adjacent edges which have different signs of SOC (clockwise
or anticlockwise along the defective geometry). Our system is in analog with
the higher-order TSCs except that the low-energy 1D model is induced by the
defect rectangle rather than the bulk topology, and the MZMs are induced by
SOC kinks rather than pairing kinks between two adjacent boundaries \cite%
{Yan2018,Wang2018}. Our system is more concise and experimentally friendly
since no tricky unconventional pairings like $d$-wave or $s_{\pm }$-wave are
required.

(iii) Our system can be realized with currently already established
experimental techniques in cold atoms, including \textrm{1D} ERD SOC by
Raman lasers \cite{Lin2011,Zhangj2012}, single-site addressing in \textrm{2D}
optical lattices \cite%
{Peter2009,Bakr2009,Bakr2010,Sherson2010,Weitenberg2011}, and tunable $s$%
-wave interaction through Feshbach resonance \cite{Thorsten2006,Chin2010}.

(iv) The Majorana corner pair is robust against shape deformation of the
defective rectangle, even to the extent of a defective loop. In addition,
the SOC direction dictates on which corners the Majorana pair resides.
Therefore the incident direction of Raman lasers can be used to manipulate
the Majorana pairs.

The paper is organized as follows. In Sec. \ref{sec2} we first introduce the
Hamiltonian with \textrm{1D} ERD SOC on optical lattices, and obtain the
phase diagram consisting of metal and $s$-wave SF phases. In Sec. \ref{sec3}%
, we study the case with a defective rectangle, and find Majorana pairs
emerge at the corners. We extend the discussion to a ring-shaped geometry in
Sec. \ref{sec4}, where the Majorana pairs arise naturally due to soft domain
walls of SOC. Finally, we make conclusions and discussions in Sec. \ref{sec5}%
.

\section{Spin-orbit-coupled $s$-wave superfluids on 2D optical lattices}

\label{sec2} 
\begin{figure}[tbp]
\centering\includegraphics[width=0.48\textwidth]{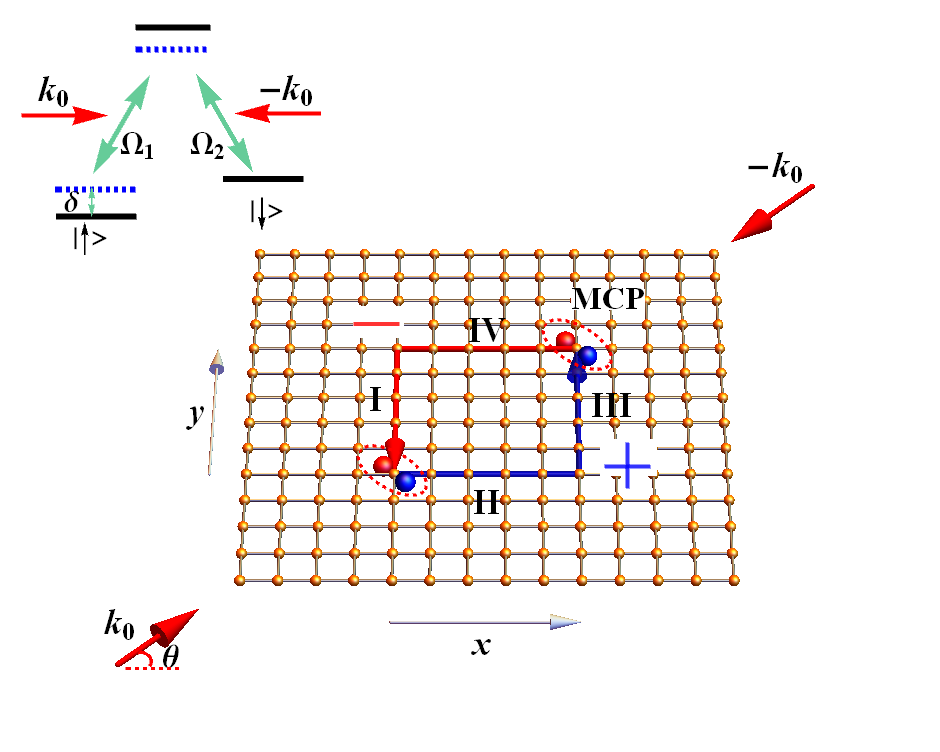}
\caption{Illustration of system setup on a 2D optical lattice, where two
counter propagating Raman beams are incident with angle $\protect\theta $. A
dip potential is applied on a rectangle geometry (indicated by the blue and
red curves) through single-site addressing. Under the configuration $\protect%
\theta =\protect\pi /4$, the sign of effective SOC is positive ($+$%
)/negative ($-$) on the blue/red lines in edge coordinate along the arrows.
Two spheres (encircled by the red dashed oval) at the interface denote the
Majorana corner pair (MCP). The inset above shows the level diagram.}
\label{Fig1}
\end{figure}

We utilize atomic hyperfine states as the pseudospin states $%
\left\vert \uparrow \right\rangle $ and $\left\vert \downarrow \right\rangle 
$, as illustrated in Fig. \ref{Fig1}. The SOC is synthesized by two
counter-propagating Raman lasers coupling the two hyperfine states. The
single atom motion in 2D real space is described by the Hamiltonian $\hat{H}%
_{0,a}=\frac{\vec{k}^{2}}{2m_{0}}+\frac{\delta }{2}\sigma _{z}+(\Omega e^{2i%
\vec{k}_{0}\cdot \vec{r}}\left\vert \downarrow \right\rangle \left\langle
\uparrow \right\vert +h.c.)$, where the reduced Planck constant $\hbar $ has
been set to be $1$, $\Omega \propto \Omega _{1}\Omega _{2}^{\ast }$ is the
strength of Raman coupling, and $\vec{k}_{0}=k_{0,x}\vec{e}_{x}+k_{0,y}\vec{e%
}_{y}$ is the wave vector of the Raman laser. The off-diagonal terms
correspond to a spin flip process accompanied by a momentum transfer of $2%
\vec{k}_{0}$, describing the SOC effects. The detuning term reads $\delta
=\omega _{z}-\delta \omega $, where $\omega _{z}>0$ is the energy difference
between these two hyperfine states, and $\delta \omega $ denotes the
frequency difference between two Raman laser beams. In the following, we
assume that other hyperfine levels are far off-resonance under the
two-phonon process, for example, by quadratic Zeeman shift. The Hamiltonian
is first transformed by a unitary matrix, namely, $\hat{H}_{0,b}=U\hat{H}%
_{0,a}U^{-1}$, where $U=$\textrm{diag}$(e^{-i\vec{k}_{0}\cdot \vec{r}},e^{i%
\vec{k}_{0}\cdot \vec{r}})$. We then perform another pseudo-spin rotations $%
\tilde{U}=e^{-i\frac{\pi }{4}\sigma _{z}}e^{-i\frac{\pi }{4}\sigma _{y}}$ to
obtain $\hat{H}_{0}=\tilde{U}\hat{H}_{0,b}\tilde{U}^{\dagger }$, which can
be written as 
\begin{equation}
\hat{H}_{0}=\frac{1}{2m_{0}}\left[ \left( k_{x}+k_{0,x}\sigma _{y}\right)
^{2}+\left( k_{y}+k_{0,y}\sigma _{y}\right) ^{2}\right] +\frac{\delta }{2}%
\sigma _{y}-\Omega \sigma _{z}.  \label{eq}
\end{equation}%
The Raman transition produces a desired ERD SOC. In the following, we assume 
$\delta =0$ for convenience.

\begin{figure}[tbp]
\centering\includegraphics[width=0.48\textwidth]{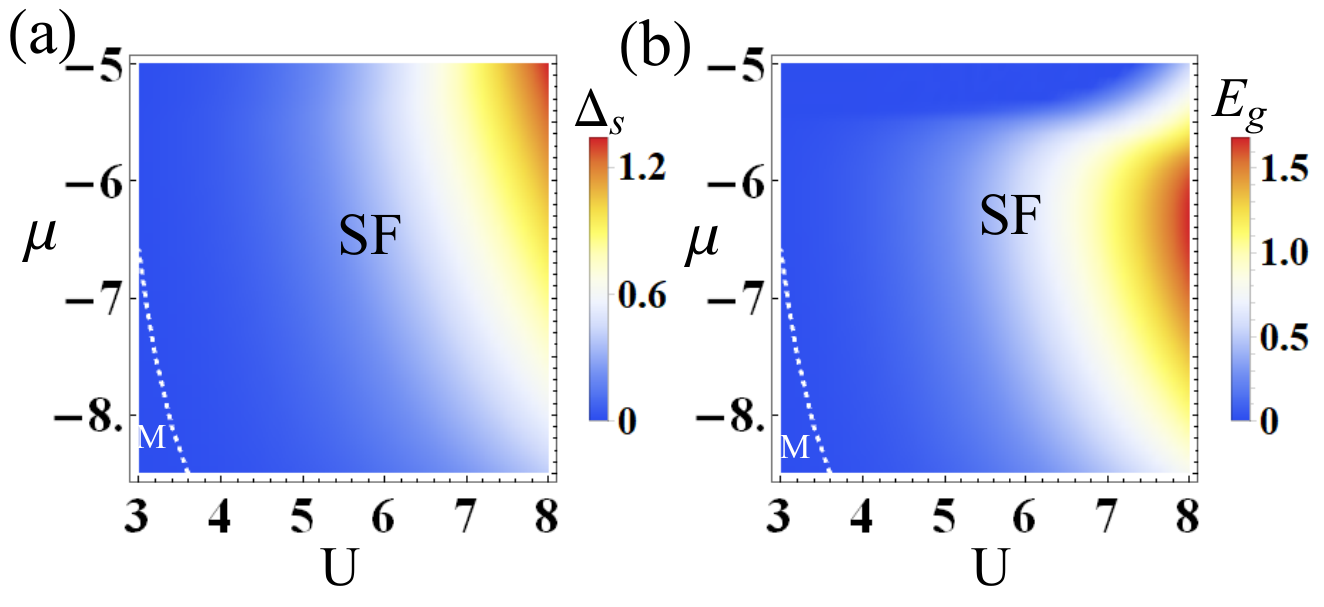}
\caption{(a) SF order parameter $\Delta _{s}$ versus interaction $U$ and
chemical potential $\protect\mu $. A phase transition occurs between metal
(M) and superfluid (SF) phases. (b) Similar as panel (a) but plotted with
Bogliubov quasiparticle energy gap $E_{g}$. In both panels, we take $%
t_{x}=t_{y}=1$, $t_{sox}=t_{soy}=2$, $h_{z}=1.4$. }
\label{Fig2}
\end{figure}

1D system suffers from strong quantum fluctuations, which could eliminate $s$%
-wave SF order, together with the Majorana modes. This motivates us to
investigate the physics in 2D, where quasi-long-range SF order exists below
the Berezinskii-Kosterlitz-Thouless (BKT) transition temperature. We
concentrate on the lowest (nearly) degenerate bands for constructing a
tight-binding model. From Eq. (\ref{eq}), through the operator $\hat{\Psi}%
\left( \vec{r}\right) =\sum_{i,\sigma }\hat{c}_{i,\sigma }\psi _{\sigma
}\left( \vec{r}-\vec{r}_{i}\right) $ with $\psi _{\sigma }\left( \vec{r}-%
\vec{r}_{i}\right) $ the Wannier function at site $i$, we obtain a
second-quantization formula: 
\begin{eqnarray}
\hat{H}_{0s} &=&\sum_{i}\left( -t_{x}\hat{c}_{i}^{\dagger }\hat{c}%
_{i+e_{x}}-t_{y}\hat{c}_{i}^{\dagger }\hat{c}_{i+e_{y}}-it_{sox}\hat{c}%
_{i}^{\dagger }\sigma _{y}\hat{c}_{i+e_{x}}\right.  \notag \\
&&\left. -it_{soy}\hat{c}_{i}^{\dagger }\sigma _{y}\hat{c}_{i+e_{y}}\right) +%
\mathrm{h.c.}-h_{z}\hat{c}_{i}^{\dagger }\sigma _{z}\hat{c}_{i}-\mu \hat{c}%
_{i}^{\dagger }\hat{c}_{i},
\end{eqnarray}%
where $t_{x}=t_{0}\cos \left( k_{0}\cos \theta \right) $, $t_{y}=t_{0}\cos
\left( k_{0}\sin \theta \right) $, $t_{sox}=t_{0}\sin \left( k_{0}\cos
\theta \right) $, $t_{soy}=t_{0}\sin \left( k_{0}\sin \theta \right) $, and
the bare hopping strength reads $t_{0}=-\int d\vec{r}\psi _{\sigma }^{\ast
}\left( \vec{r}-\vec{r}_{i}\right) \left( \frac{k_{x}^{2}+k_{y}^{2}}{2m_{0}}%
+V_{\mathrm{lat}}\right) \psi _{\sigma }\left( \vec{r}-\vec{r}_{i+1}\right) $%
. Here, we have chosen the basis $\hat{c}_{i}=\left( \hat{c}_{i,\uparrow },%
\hat{c}_{i,\downarrow }\right) ^{T}$ and denoted $h_{z}=\Omega $. The
incident angle $\theta $ is illustrated in Fig.~\ref{Fig1}. The lattice
spacing is set to be $a=1$. Hereafter, we set $t_{0}=1/\cos \left[ k_{0}\sin
\left( \pi /4\right) \right] =1/\cos (\sqrt{2}\pi /4)$ for convenience. When 
$\theta =\pi /4$, we have approximately $t_{sox}=t_{soy}\approx
2t_{x}=2t_{y} $.

We consider an attractive SU$(2)$-invariant interaction $\hat{H}%
_{int}=-\sum_{i}U\hat{n}_{i,\uparrow }\hat{n}_{i,\downarrow }$ and study the
superfluid phase under mean-field approach by solving the $s$-wave
superfluid order parameter $\Delta _{s}=U\left\langle c_{i,\downarrow
}c_{i,\uparrow }\right\rangle $ self-consistently. The Bogoliubov de Genns
(BdG) Hamiltonian in the Nambu basis $\Psi _{k}=\left( c_{k\uparrow
},c_{k\downarrow },c_{-k\downarrow }^{\dagger },-c_{-k\uparrow }^{\dagger
}\right) ^{T}$ is described by $\hat{H}_{s}=\sum_{k}\Psi _{k}^{\dagger
}H\left( k\right) \Psi _{k}$ with 
\begin{equation}
H\left( k\right) =\left( \epsilon _{k}+\gamma _{k}\sigma _{y}\right) \tau
_{z}-h_{z}\sigma _{z}+\Delta _{s}\tau _{x},  \label{eq0}
\end{equation}%
where $\epsilon _{k}=-2\left( t_{x}\cos k_{x}+t_{y}\cos k_{y}\right) -\mu $, 
$\gamma _{k}=2\left( t_{sox}\sin k_{x}+t_{soy}\sin k_{y}\right) $, $\sigma $
and $\tau $ are Pauli matrices acting on the spin and particle-hole spaces,
respectively. By minimizing free energy with respect to the order parameter $%
\Delta _{s}$ and chemical potential $\mu $, we may derive the following
self-consistent equations 
\begin{eqnarray}
1 &=&\frac{U}{2N_{l}}\sum_{\nu =\pm ,k}\frac{\tanh \left( \beta \xi _{k,\nu
}/2\right) }{\xi _{k,\nu }}\left( 1+\frac{\nu h_{z}^{2}}{g_{k}}\right) ,
\label{ef1} \\
n_{f} &=&1-\frac{1}{N_{l}}\sum_{\nu =\pm ,k}\epsilon _{k}\frac{\tanh \left(
\beta \xi _{k,\nu ,}/2\right) }{\xi _{k,\nu }}\left( 1+\frac{\nu m_{k}^{2}}{%
g_{k}}\right)  \label{ef2}
\end{eqnarray}%
where $n_{f}$ is the particle filling factor, $N_{l}$ is the number of
lattice sites, and other parameters are defined as $\beta =1/\left(
k_{B}T\right) $ with $k_{B}$ the Boltzmann constant and $T$ the temperature, 
$\xi _{k,\nu =\pm }=\sqrt{\epsilon _{k}^{2}+\Delta _{s}^{2}+m_{k}^{2}+2\nu
g_{k}}$, $g_{k}=\sqrt{\epsilon _{k}^{2}m_{k}^{2}+\Delta _{s}^{2}h_{z}^{2}}$,
and $m_{k}=\sqrt{h_{z}^{2}+\left\vert \gamma _{k}\right\vert ^{2}}$. By
numerically solving equations (\ref{ef1}) and (\ref{ef2}), we obtain phase
diagrams at zero temperature for pairing order $\Delta _{s}$ and
quasiparticle energy gap $E_{g}$ in Fig. \ref{Fig2} (a) and (b),
respectively. Fig. \ref{Fig2} (a) confirms the phase transition from a metal
(M) phase to an $s$-wave SF. From panel (b), we find a finite gap for
Bogoliubov quasiparticle excitations in proper parameter region in the SF
phase. The energy gap also survives on a finite-size sample and could
protect Majorana modes from lower extended states.

\section{Majorana corner pairs on a topological defective rectangular
geometry}

\label{sec3} Given a proper local dip potential, the defect chain enjoys a
non-trivial topology, belonging to the BDI class. It can be characterized
through a winding number, which is discussed in Appendix \ref{App1}.
Similarly, with a local dip $\mu _{d}$, we can get a defect rectangle in the
2D optical lattice as illustrated in Fig. \ref{Fig1}, where SOC domain walls
(anticlockwise or clockwise) naturally arise at two corners. In the
following, we will first focus on the continuum limit to explore the nature
of the emerged Majorana pairs, supplemented with self-consistent numerical
calculations on a 2D optical lattice.

We assume that with appropriate $\mu _{d}$, the defect rectangle enters the
TSF phase while the rest part remains trivial. As a result, we could assume
that the topological defect rectangle is isolated from the \textrm{2D} bulk.
The numerics performed on a \textrm{2D} optical lattice with an imprinted
defect-rectangle also supports this assumption later. From Eq. (\ref{eq0}),
the low-energy Hamiltonian expands around $\vec{k}=\left( 0,0\right) $ on
edges $\mathrm{m}=$\textrm{I}, \textrm{II}, \textrm{III}, \textrm{IV} (see
Fig. \ref{Fig1}) and is then given by%
\begin{equation}
H_{\mathrm{m}}=t_{\mathrm{m}}k_{\mathrm{m}}^{2}\tau _{z}+2t_{s,\mathrm{m}}k_{%
\mathrm{m}}\sigma _{y}\tau _{z}-\mu _{\mathrm{m}}\tau _{z}-h_{z}\sigma
_{z}+\Delta _{s,\mathrm{m}}\tau _{x},
\end{equation}%
where $t_{\mathrm{I}}=t_{\mathrm{III}}=t_{y}$, $t_{\mathrm{II}}=t_{\mathrm{IV%
}}=t_{x}$, $k_{\mathrm{I}}=k_{\mathrm{III}}=k_{y}$, $k_{\mathrm{II}}=k_{%
\mathrm{IV}}=k_{x}$, $t_{s,\mathrm{I}}=t_{s,\mathrm{III}}=2t_{soy}$ and $%
t_{s,\mathrm{II}}=t_{s,\mathrm{IV}}=2t_{sox}$. The on-site chemical
potential is $\mu _{\mathrm{m}}=\mu +2\left( t_{x}+t_{y}\right) -\mu _{d}$
with $\mu _{d}$ the dip potential, and $\Delta _{s,\mathrm{m}}$ the $s$-wave
pairing on each edge. Without loss of generality, we set incident angle of
Raman lasers $\theta =\frac{\pi }{4}$ such that $t_{x}=t_{y}=t$, $%
t_{sox}=t_{soy}=t_{\mathrm{so}}$, $\mu _{\mathrm{m}}=\mu _{\mathrm{edge}%
}=\mu +4t-\mu _{d}$, and assume the $s$-wave SF order parameter is nearly
uniform on the four edges $\Delta _{s,\mathrm{m}}=\Delta _{\mathrm{edge}}$.
For later convenience, we take an \textquotedblleft edge coordinate" $s$, in
which we take the anticlockwise direction as positive. In such a coordinate,
the low-energy edge Hamiltonian reads%
\begin{equation}
H_{\mathrm{edge}}=-t\tau _{z}\frac{\partial ^{2}}{\partial s^{2}}-i\alpha
\left( s\right) \sigma _{y}\tau _{z}\frac{\partial }{\partial s}-\mu _{%
\mathrm{edge}}\tau _{z}-h_{z}\sigma _{z}+\Delta _{\mathrm{edge}}\tau _{x},
\label{eq2D}
\end{equation}%
with $\alpha \left( s\right) =-2t_{so}$, $2t_{so}$, $2t_{so}$, $-2t_{so}$
for edge \textrm{I}-\textrm{IV} respectively. Remarkably, while the terms $%
\Delta _{\mathrm{edge}}$ and $\mu _{\mathrm{edge}}$ remain the same on the
four edges, the effective coupling $\alpha \left( s\right) $ changes sign at
two of four corners (the corner between the edges \textrm{I (III)} and 
\textrm{II (IV)}), forming two SOC domain walls as illustrated in Fig. \ref%
{Fig1}. This will give rise to a Majorana pair if $h_{z}^{2}>\mu _{\mathrm{%
edge}}^{2}+\Delta _{\mathrm{edge}}^{2}$. Specifically, at the corner between
edge \textrm{I} and \textrm{II} (corner $s=0$ in our coordinate), two
orthogonal wave functions for MCMs are given by%
\begin{equation}
\Psi _{0,\pm }=C_{\pm }e^{-\eta _{\pm }\left\vert s\right\vert }\left( e^{i%
\frac{\phi _{\pm }}{2}}\left\vert y_{+}\right\rangle _{\sigma }\left\vert
y_{\pm }\right\rangle _{\tau }+e^{-i\frac{\phi _{\pm }}{2}}\left\vert
y_{-}\right\rangle _{\sigma }\left\vert y_{\mp }\right\rangle _{\tau
}\right) .  \label{eq2DS}
\end{equation}%
Here, $C_{\pm }$ are normalization constants and $e^{i\phi _{\pm }}=\frac{%
h_{z}\left[ i\left( \alpha \eta _{\pm }\mp \Delta _{\mathrm{edge}}\right)
-\left( t\eta _{\pm }^{2}+\mu _{\mathrm{edge}}\right) \right] }{\left( t\eta
_{\pm }^{2}+\mu _{\mathrm{edge}}\right) ^{2}+\left( \alpha \eta _{\pm }\mp
\Delta _{\mathrm{edge}}\right) ^{2}}$, where $\eta _{\pm }=\frac{1}{2}\sqrt{-%
\frac{2\varkappa }{3}+\delta }\mp \frac{1}{2}\sqrt{-\frac{4\varkappa }{3a}%
-\delta +\zeta _{\pm }}>0$, $\zeta _{\pm }=\mp 2d/(a\sqrt{-2c/(3a)+\delta }$%
, $\delta =\frac{\sqrt[3]{2}\delta _{1}}{3a\sqrt[3]{\delta _{2}+\sqrt{%
-4\delta _{1}^{3}+\delta _{2}^{2}}}}+\frac{\sqrt[3]{\delta _{2}+\sqrt{%
-4\delta _{1}^{3}+\delta _{2}^{2}}}}{3\sqrt[3]{2a}}$, $\delta _{1}=\varkappa
^{2}+12ae$, $\delta _{2}=2\varkappa ^{3}+27ad^{2}-72a\varkappa e$, $a=t^{2}$%
, $\varkappa =\alpha ^{2}+2t\mu _{\mathrm{edge}}$, $d=-2\alpha \Delta _{%
\mathrm{edge}}$ and $e=\Delta _{\mathrm{edge}}^{2}-h_{z}^{2}+\mu _{\mathrm{%
edge}}^{2}$. The vectors $\left\vert y_{\pm }\right\rangle _{\sigma }$ and $%
\left\vert y_{\pm }\right\rangle _{\tau }$ are eigenstates of operators $%
\sigma _{y}$ and $\tau _{y}$, respectively. Following similar approach, we
could also find two Majorana modes at the corner between edges \textrm{III}
and \textrm{IV} (see Appendix \ref{App2} for details). We emphasize that as
long as the four edges are in the TSF phase, the very existence of Majorana
pairs is robust against the fluctuations of chemical potential and SF order
parameter.

\begin{figure}[tbp]
\centering\includegraphics[width=0.48\textwidth]{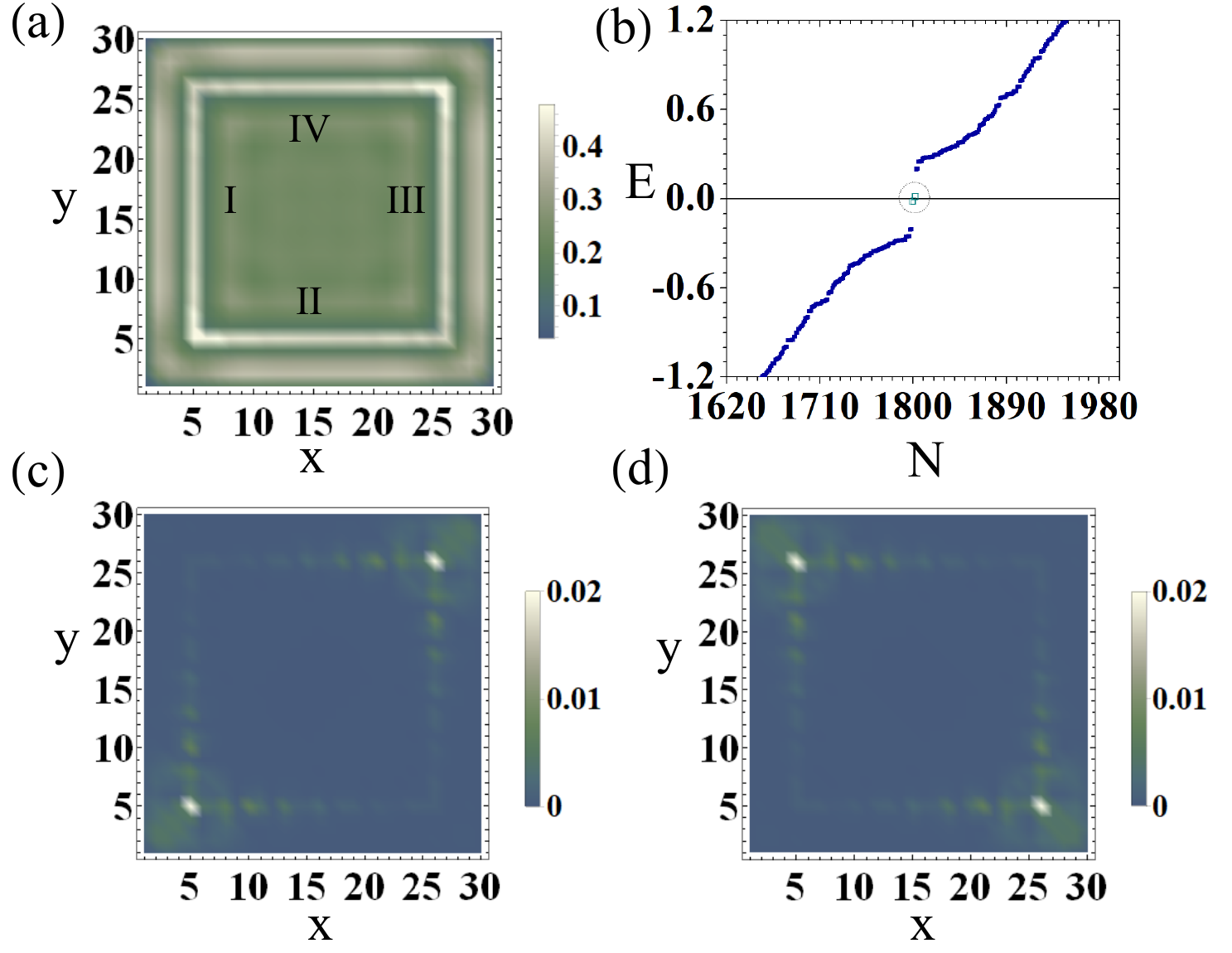}
\caption{(a) Self-consistent $s$-wave pairing order parameters in real
space. (b) Energy level diagrams. (c,d) The real-space distributions of
MZMs. For all panels, we choose $t_{x}=t_{y}=1,t_{sox}=t_{soy}=2$, $\protect%
\mu =-7.95$, $\protect\mu _{d}=-4.0$, $h=1.4$, $U=6.3$. The incident angle
is $\protect\theta =\protect\pi /4$ in (a)-(c) and $\protect\theta =-\protect%
\pi /4$ in (d).}
\label{Fig3}
\end{figure}

With the above understanding of continuum systems, we now proceed to study
the discrete cases on an optical lattice shown in Fig. \ref{Fig1}. The total
Hamiltonian now becomes 
\begin{equation}
\hat{H}_{BdG}=\hat{H}_{s}+\sum_{i\in \square }\mu _{d}\hat{c}_{i}^{\dagger }%
\hat{c}_{i},
\end{equation}%
where $i\in \square $ enumerate each site with the dip potential (a
rectangular geometry in this case). The local $s$-wave superfluid order
parameter in real space is determined in a self-consistent manner \cite%
{Jiang2016}, as well as the quasiparticle energy spectra and wave functions.
On the defect rectangle, the system is topological once $h_{z}>\sqrt{\tilde{%
\mu}^{2}+\Delta _{s}^{2}}$ and $\tilde{\mu}=\mu +2t_{x}+2t_{y}-\mu _{d}$. In
our self-consistent numerical calculations, we take the lattice sizes $%
n_{x}=n_{y}=30$, and the defect rectangle is given by $%
n_{x}^{d}=n_{y}^{d}=22 $. The SF order parameter $\Delta _{s,i}$ is shown in
Fig. \ref{Fig3} (a), which has a constant phase across the entire system.
Fig. \ref{Fig3} (b) shows the quasiparticle energy spectrum, where four
Majorana bound states (two Majorana corner pairs) exist in the energy gap. A
small energy splitting is observed as a result of finite-size effect. Fig. %
\ref{Fig3} (c) shows the density distribution of the bounded Majorana corner
states, which clearly demonstrates its localization at the corners of the
defect rectangle. The Majorana corner pairs are robust against the
perturbations of chemical potential and SF order parameter that preserve
chiral symmetry. We have confirmed this point by numerical calculations.

\begin{figure}[tbp]
\centering\includegraphics[width=0.48\textwidth]{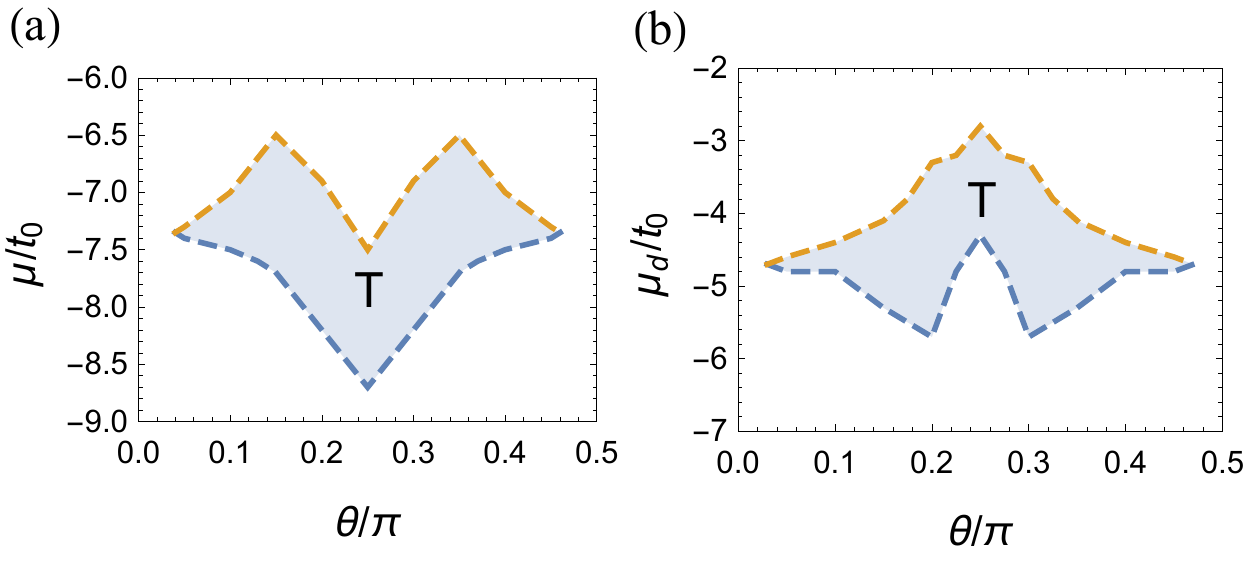}
\caption{Phase diagram for 2D system with the defect rectangle. In region
\textquotedblleft T\textquotedblright , Majorana bound states exist at
corners. In (a), $\protect\mu _{d}=-4.0$, $h_{z}=1.4$, $U=6.3$ are used. In
(b), $\protect\mu =-7.8$, $h_{z}=1.4$, $U=6.3$ are used. }
\label{Fig4}
\end{figure}

The incident angle of Raman lasers can change the SOC and the
nearest-neighbor hopping, and thus alter the Majorana bound states. Figs.~%
\ref{Fig4} (a) and (b) illustrate the corresponding phase diagram with
respect to $\mu $-$\theta $ and $\mu _{d}$-$\theta $, where Majorana corner
pairs exist in the topological region (T). It is found that the Majorana
pairs is also robust to certain variation of the incident angle $\theta $.
We remark that if the sign of $\theta $ is reversed, the Majorana pairs
appear at another two corners (the interfaces of \textrm{II-III }and\textrm{%
\ I-IV}) as shown in Fig.~\ref{Fig3}(d), which can be compared with Fig.~\ref%
{Fig3}(c). Thus, our proposed setup provides better tunability for
manipulating Majorana bound states.

\section{Majorana corner pairs on a ring geometry}

\label{sec4} 
\begin{figure}[tbp]
\centering\includegraphics[width=0.48\textwidth]{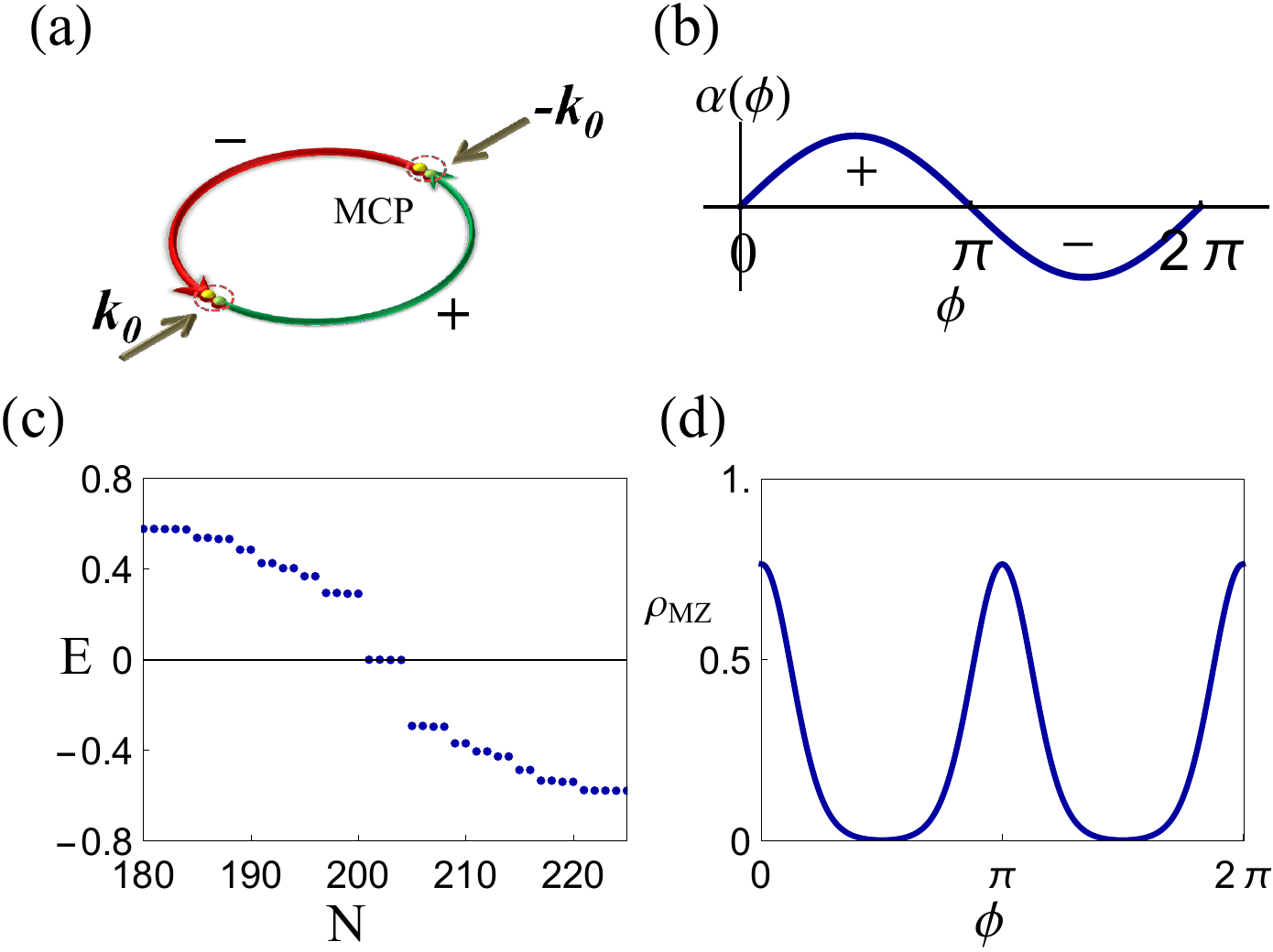}
\caption{ (a) Illustration of a ring trap in which the atoms are confined.
Two counter-propagating Raman lasers with wavevector $\pm \vec{k}_{0}$
couple the atomic hyperfine states. The notations $+$ and $-$ indicate the
sign of SOC. Two spheres encircled by the red dashed circle denote a MCP.
(b) The SOC term $\protect\alpha \left( \protect\phi \right) $ plotted to
varying polar angle $\protect\phi $. (c) The eigenspectrum computed through
plane-wave expansion. (d) The particle density $\protect\rho _{_{MZ}}$ of
Majorana zero modes versus the angle $\protect\phi $. The parameters are
chosen as $\tilde{\protect\alpha}_{0}=0.04$, $\protect\eta =0.005,\protect%
\mu ^{\prime }=0.2,h_{z}=-1.4,\Delta _{s}=0.8$. }
\label{Fig5}
\end{figure}

In this section, we study the case with a ring-shaped defect
line. Here, the optical lattice is removed and we focus on the low-energy
effective 1D model for simplicity. The effective model is illustrated in
Fig. \ref{Fig5} (a) and we find that soft domain walls of SOC naturally
arise on the ring, which leads to the emergence of Majorana corner pairs.

Without loss of generality, we assume the momentum kick by Raman lasers is
along the $x$ direction. Under a spin-rotation $\sigma _{y}\rightarrow $ $%
\sigma _{x}$ with $k_{0,y}=0$ in Eq. (\ref{eq}), the SOC has the form $%
\alpha _{0}k_{x}\sigma _{x}$, where the coupling constant is given by the
ratio of laser wavevector and atomic mass, i.e., $\alpha _{0}=k_{0}/m_{0}$.
Hence, in the continuum limit, the effective Hamiltonian reads%
\begin{equation}
\mathcal{H}=\frac{k^{2}}{2m_{0}}\tau _{z}+\alpha _{0}k_{x}\sigma _{x}\tau
_{z}-\mu \tau _{z}+h_{z}\sigma _{z}+\Delta _{s}\tau _{x},  \label{eq1a}
\end{equation}%
where $\Delta _{s}$ is an $s$-wave SF order. For simplicity, we set $\Delta
_{s}$ to be real. The relation $h_{z}>\sqrt{\mu ^{2}+\Delta _{s}^{2}}$ holds
in the topological regions.

In a polar coordinate $(\rho ,\phi )$, the above Hamiltonian becomes a
function of polar angle $\phi $ on a ring with given radii $\rho $, i.e., 
\begin{equation}
\mathcal{H}\left( \phi \right) \mathcal{=}-\eta \partial _{\phi }^{2}\tau
_{z}+i\tilde{\alpha}_{0}\sin \phi \frac{\partial }{\partial \phi }\sigma
_{x}\tau _{z}-\mu ^{\prime }\tau _{z}+h_{z}\sigma _{z}+\Delta _{s}\tau _{x},
\label{eq2}
\end{equation}%
where $\eta =\frac{1}{2m_{0}\rho ^{2}}$, $\tilde{\alpha}_{0}=\frac{\alpha
_{0}}{\rho }$ and $\mu ^{\prime }=\mu -\frac{k_{0}^{2}}{2m_{0}}$ (more
details are discussed in Appendix \ref{App3}). The Hamiltonian $\mathcal{H}%
\left( \phi \right) $ has particle-hole symmetry $\mathcal{PH}\left( \phi
\right) \mathcal{P}^{-1}=-\mathcal{H}\left( -\phi \right) $ where $\mathcal{P%
}=\sigma _{y}\tau _{y}\mathcal{K}$ and $\mathcal{K}$ denotes the complex
conjugation. It also preserves a generalized time-reversal symmetry $%
\mathcal{TH}\left( \phi \right) \mathcal{T}^{-1}=\mathcal{H}\left( -\phi
\right) $ with $\mathcal{T}=\sigma _{z}\mathcal{K}$. The combination of $%
\mathcal{P}$ and $\mathcal{T}$ leads to the chiral symmetry $\mathcal{C}$: $%
\mathcal{CH}\left( \phi \right) \mathcal{C}^{-1}=-\mathcal{H}\left( \phi
\right) $, with $\mathcal{C}=\mathcal{PT}=i\sigma _{x}\tau _{y}$. Therefore,
the Hamiltonian belongs to $\mathrm{BDI}$ class and can be characterized by
a $\mathbb{Z}$ topological invariant \cite{Altland1997,Schnyder2008}.

In Eq. (\ref{eq2}), the SOC $\alpha \left( \phi \right) =\tilde{\alpha}%
_{0}\sin \phi $\ changes sign at \ $\phi =0,\pi $, as shown in Fig. \ref%
{Fig5} (b). Specifically, we have $\alpha \left( \phi \right) >$\ $0$ if $%
\phi \in \left( 0,\pi \right) $ and $\alpha \left( \phi \right) <0$ if $\phi
\in \left( \pi ,2\pi \right) $. Hence, the system can be divided into two
segments. Both belong to the $\mathrm{BDI}$\ class but possess opposite
topological invariant. The interfaces are determined by $\phi =0$ and $\pi $%
, corresponding to two \textquotedblleft soft\textquotedblright\ domain
walls in the sense that the SOC term changes smoothly across these two
points. From Eq. (\ref{eq2}), the Hamiltonian $\mathcal{H}\left( \phi
\right) $ is invariant under a $2n\pi $\ rotation if $n$ is an integer.
Therefore, to solve the eigenvalues of $\mathcal{H}\left( \phi \right) $, we
assume the following trial solution, 
\begin{equation}
\Phi \left( \phi \right) =\left( 
\begin{array}{cccc}
u_{a}\left( \phi \right) , & u_{b}\left( \phi \right) , & u_{c}\left( \phi
\right) , & u_{d}\left( \phi \right)%
\end{array}%
\right) ^{T},
\end{equation}%
where $u_{\nu =a,b,c,d}\left( \phi \right) =\sum_{m}\nu _{m}e^{im\phi }$ and 
$m$ is an integer. By solving the Schr\"{o}dinger equation $\mathcal{H}%
\left( \phi \right) \Phi \left( \phi \right) =E\Phi \left( \phi \right) $,
the eigenvalues are obtained as shown in Fig. \ref{Fig5} (c). See Appendix %
\ref{App3} for more details. It is clear that four Majorana modes emerge
(with an numerical error about $E\approx 10^{-4}$). One Majorana corner pair
consisting of two Majorana modes localizes at $\phi =0$, and the other pair
localizes at $\phi =\pi $, as illustrated in Fig. \ref{Fig5} (a). This is
also demonstrated by the particle density distribution $\rho _{_{MZ}}$ of
Majorana modes, as shown in Fig. \ref{Fig5} (d). We remark that a toroidal
Bose-Einstein condensate has been created in an all-optical trap \cite%
{Ramanathan2011}. We expect our scheme could be reached with similar
techniques and additional Raman lasers.

\section{Discussion and Conclusion}

\label{sec5} From the effective low-energy theory of TSFs, it is well-known
that Majorana modes would emerge if the sign of the Dirac mass changes and
most previous proposals are based on this principle. In this paper, we
propose an alternative approach to implement Majorana modes (Majorana corner
pairs) through tuning the effective SOC. By loading Fermi gases on \textrm{2D%
} optical lattices subjected to a \textrm{1D} ERD SOC, we can find a SF
phase under appropriate $s$-wave interaction and Zeeman field. Using
single-site addressing techniques, we could engineer defective geometries,
which are topologically non-trivial, on the 2D optical lattice. From the
viewpoint of low-energy theories, a defect rectangle consists of two TSFs
characterized by distinct topological invariants whose sign is determined by
the sign of SOC in edge coordinate. Obviously, the sign of SOC changes at
two corners on the defect rectangle. At the interface of two distinct TSF, a
topologically protected Majorana pair naturally arises according to the
index theorem. For TSF with \textrm{1D} ERD SOC on a ring, two soft SOC
domain walls exist, and two Majorana pairs also appear near the domain
walls. In principle, as long as two effective \textrm{1D} SFs are
topological with different topological invariants $w=\pm 1$, the Majorana
pair will emerge at the interface. It is robust as long as the perturbations
preserve three underlying symmetries ($\mathcal{P}$, $\mathcal{T}$, $%
\mathcal{C}$) of the system.

We emphasize that the Majorana corner pair in the context differs from those
in second order TSCs in two dimensions. First, for second order TSCs with
time reversal symmetry \cite{Yan2018,Wang2018}, \textrm{1D} edge modes
evolve from the higher-dimensional bulk of the topological insulators.
However, in our scheme, the \textrm{1D} modes originate from the defect
geometry. Second, in a higher-order TSC, a momentum-dependent SC pairing ($%
s_{\pm }$ or $d$-wave) leads to Dirac mass kink at the corner of the sample,
and then induces the Majorana Kramers pair. In contrary, our proposal
utilizes the sign reverse of effective SOC on the edges and lacks Kramers
degeneracy.

In summary, we propose a distinct scheme to implement Majorana pairs in an
atomic platform. The coordinate of Majorana pair depends on the position of
SOC domain wall which can be tuned by the directions of the Raman laser
beams. Moreover, our system is free of dynamical instability such that the
MZM has a longer lifetime. Our work opens the possibility of implementing
robust Majorana pairs and the associated non-Abelian braiding in cold atoms.

\begin{acknowledgments}
This work is supported by Air Force Office of Scientific Research
(FA9550-16-1-0387), National Science Foundation (PHY-1806227), and Army
Research Office (W911NF-17-1-0128). This work is also supported in part by
NSFC under the grant No. 11504285, and the Scientific Research Program
Funded by Natural Science Basic Research Plan in Shaanxi Province of China
(Program Nos. 2018JQ1058), the Scientific Research Program Funded by Shaanxi
Provincial Education Department under the grant No. 18JK0397, and the
scholarship from China Scholarship Council (CSC) (Program No. 201708615072).
\end{acknowledgments}

\twocolumngrid

\appendix

\section{Majorana modes at the ends of topological defect-chain}

\label{App1} In this section, we show a topologically non-trivial defective
chain can be implemented through on-site potential engineering on \textrm{2D}
optical lattices.

For a 1D system with SOC and SF order, the system is topological if $h_{z}>%
\sqrt{\left( \tilde{\mu}-\mu _{d}\right) ^{2}+\Delta _{s}^{2}}$ , where $%
\tilde{\mu}=\mu +2t_{x}+2t_{y}$ \cite%
{Lutchyn2010,Oppen2010,Jiang2016,Sau2010}. Reference \cite{Jiang2016} shows
Majorana fermions may be generated in a 2D optical lattices with 1D ERD SOC
along the $x$ direction. This motivates us to demonstrate the existence of
Majorana bound states in a genuine 2D systems with 1D defects, where the SOC
lays along $\vec{e}_{x}+\vec{e}_{y}$ direction. Through single-site
addressing, a potential could be locally applied to a given site.

\begin{figure}[tbp]
\centering\includegraphics[width=0.48\textwidth]{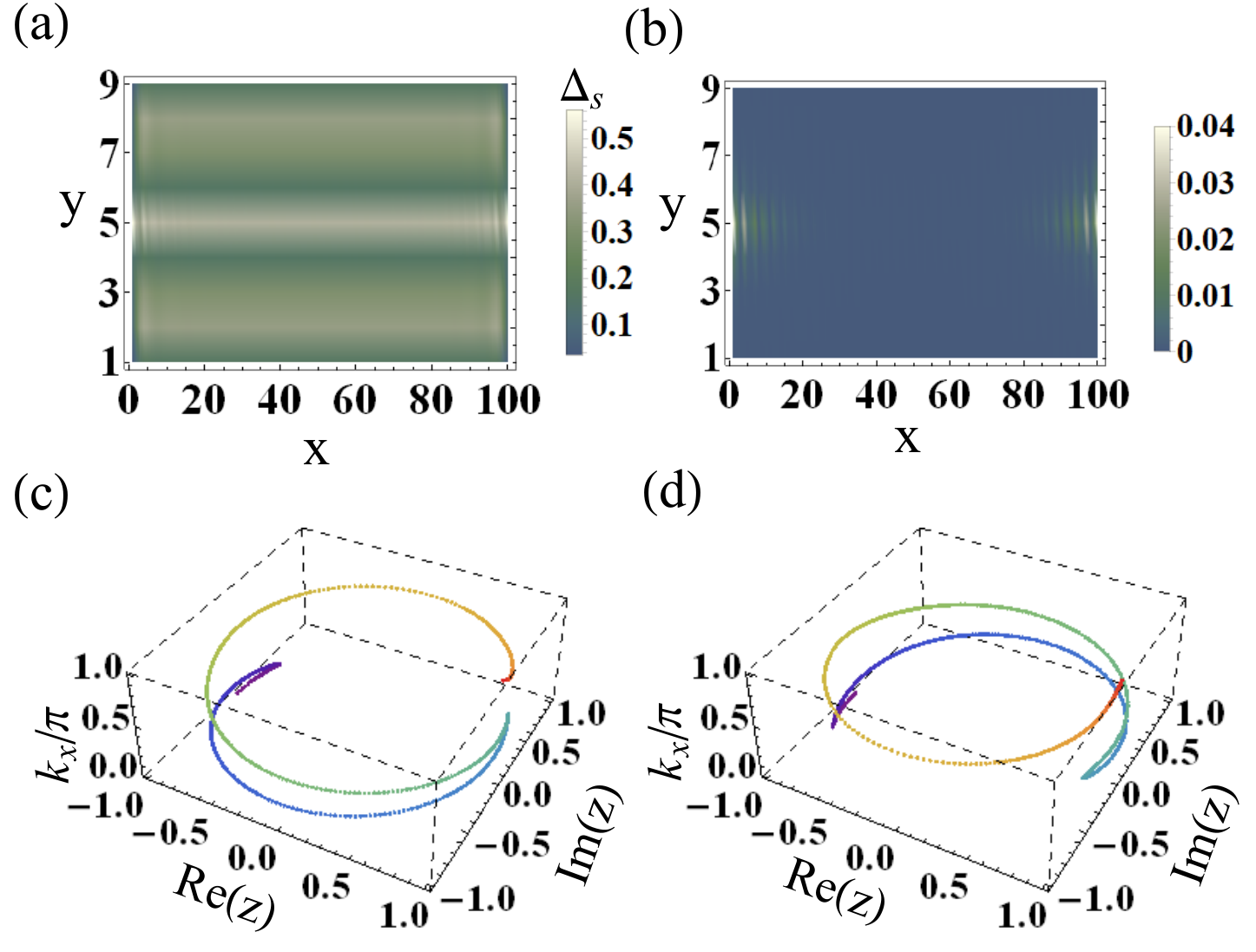}
\caption{\ (a) The amplitude of SF order parameter in real space. (b) The
density distribution of the MZM. (c,d) $z(k_{x})$ in the complex plane for
opposite winding number. In all panels, we set $t_{x}=t_{y}=1$, $h_{z}=1.4$, 
$\protect\mu =-8.0$, $\protect\mu _{d}=-4$, $\Delta _{s}=0.3$ and $\Delta
_{a}=0.11$. We choose $t_{sox}=t_{soy}=2$ in (a-c) and $t_{sox}=t_{soy}=-2$
in (d).}
\label{Fig6}
\end{figure}

Imposing a 1D potential dip $\mu _{d}$, we have the following Hamiltonian 
\begin{equation}
\hat{H}_{BdG}=\hat{H}_{s}+\sum_{i\in -}\mu _{d}\hat{c}_{i}^{\dagger }\hat{c}%
_{i},
\end{equation}%
where $i\in -$ denotes the sites $(i_{x},i_{y})$ satisfying $i_{y}=n_{y_{c}}$%
. In self-consistent numerical calculations, we take a lattice with $%
n_{x}=100$, $n_{y}=9$ and $n_{y_{c}}=5$ under open boundary conditions,
where $n_{x}$ and $n_{y}$ denote the site number along the $x$ and $y$
directions. Figs. \ref{Fig6} (a) and (b) present self-consistent \ numerical
results. Fig. \ref{Fig6} (b) shows density profile of the zero-energy mode ($%
E\approx 10^{-4}$). It demonstrates the existence of MZMs even in a genuine 
\textrm{2D} system. From self-consistent BdG numerical results, the SF order
parameter is almost homogeneous along the $x$ direction as shown in Fig. \ref%
{Fig6} (a). Thus, with periodic boundary condition, the system has a
translation symmetry along the $x$ direction so that momentum $k_{x}$ is a
good quantum number. The 2D optical lattice can be regarded as layered 
\textrm{1D} chain with transverse tunneling and SOC effects. The effective
Hamiltonian is then written as%
\begin{eqnarray}
H_{BdG}(k_{x}) &=&\kappa _{0}h_{0}\left( k_{x}\right) +\mu _{d}\kappa
_{c}\sigma _{0}\tau _{z}+\Delta _{a}\kappa _{c}\sigma _{y}\tau _{y}  \notag
\\
&&-t_{y}\kappa _{x}\sigma _{0}\tau _{z}-t_{soy}\kappa _{y}\sigma _{y}\tau
_{z}.  \label{BdG}
\end{eqnarray}%
Here, the matrix $\kappa $ acts on chain space, with $\kappa _{0}$ identity
matrix, $\left( \kappa _{c}\right) _{i,i}=1$ for $i=n_{y_{c}}$ and $0$
otherwise, $\left( \kappa _{x}\right) _{i,j}=1$ for $\left\vert
i-j\right\vert =1$ and $\left( \kappa _{y}\right) _{i,i\mp 1}=\mp i$ and $0$
otherwise. The term proportional to $\mu _{d}$ ($\Delta _{a}$) describes the
dip potential (the SF-order) difference between the central chain and other
individual chains. The term proportional to $t_{y}$ ($t_{soy}$) describes
the hopping (SOC) along the $y$ direction. The following Hamiltonian 
\begin{eqnarray}
h_{0}\left( k_{x}\right) &=&\left( -2t_{x}\cos k_{x}-\mu \right) \sigma
_{0}\tau _{z}+2t_{sox}\sin k_{x}\sigma _{y}\tau _{z}  \notag \\
&&+h_{z}\sigma _{z}\tau _{z}+\Delta _{s}\sigma _{y}\tau _{y}
\end{eqnarray}%
describes the original uniform individual 1D chain along the $x$ direction. $%
\sigma $ and $\tau $ are Pauli matrices acting on spin space and
particle-hole space, respectively. The above BDG Hamiltonian $H_{BdG}(k_{x})$
has intrinsic particle-hole symmetry $\mathcal{P}$: $\mathcal{P}%
H_{BdG}\left( k_{x}\right) \mathcal{P}^{-1}=-H_{BdG}\left( -k_{x}\right) $
with $\mathcal{P}=\tilde{\tau}_{x}\mathcal{K}$, $\tilde{\tau}_{x}=\tau
_{x}\sigma _{0}\eta _{0}$, where $\sigma _{0}$ is $2\times 2$ identity
matrix, $\eta _{0}$ is a $N_{s}\times N_{s}$ identity matrix acting on the
lattice site space, and $\mathcal{K}$ is the complex conjugation. If the
superfluid order parameter is real (or has a constant phase that can be
eliminated by gauge transformations), the Hamiltonian preserves a
generalized time-reversal symmetry $\mathcal{T}$: $\mathcal{T}H_{BdG}\left(
k_{x}\right) \mathcal{T}^{-1}=H_{BdG}\left( -k_{x}\right) $ with $\mathcal{T}%
=\mathcal{K}$. The composite operation of $\mathcal{P}$ and $\mathcal{T}$
also leads to a chiral symmetry $\mathcal{C}$: $\mathcal{C}H_{BdG}\left(
k_{x}\right) \mathcal{C}^{-1}=-H_{BdG}\left( k_{x}\right) $, with $\mathcal{C%
}=\mathcal{P}\mathcal{T}=\tilde{\tau}_{x}$. From above symmetry analyses,
the Hamiltonian belongs to $\mathrm{BDI}$ class, characterized by a $\mathbb{%
Z}$ topological invariant (winding number). 

The winding number $w$ can characterize the topological properties of BdG
Hamiltonian (\ref{BdG}) \cite{Tewari2012}. Because the BdG Hamiltonian $%
H_{BDG}$ has the chiral symmetry, it can be transformed into an off-diagonal
form in particle-hole space under a unitary transformation $U=e^{-i\frac{\pi 
}{4}\tau _{y}}$,%
\begin{equation}
UH_{BdG}(k_{x})U^{-1}=\left( 
\begin{array}{cc}
0 & \mathcal{B}\left( k_{x}\right) \\ 
\mathcal{B}^{T}\left( -k_{x}\right) & 0%
\end{array}%
\right) .
\end{equation}%
Here, $\mathcal{B}\left( k_{x}\right) =\mathcal{B}_{1}\left( k_{x}\right) -i%
\mathcal{B}_{2}$, where $\mathcal{B}_{1}\left( k_{x}\right) =\left(
-2t_{x}\cos k_{x}-\mu \right) \kappa _{0}\sigma _{0}+2t_{sox}\sin
k_{x}\kappa _{0}\sigma _{y}+h_{z}\kappa _{0}\sigma _{z}+\mu _{d}\kappa
_{c}\sigma _{0}-t_{y}\kappa _{x}\sigma _{0}-t_{soy}\kappa _{y}\sigma _{y}$
and $\mathcal{B}_{2}=\left( \Delta _{s}\kappa _{0}\sigma _{y}+\Delta
_{a}\kappa _{c}\sigma _{y}\right) $. The winding number is defined as \cite%
{Tewari2012} 
\begin{equation}
w=-\frac{i}{\pi }\int_{0}^{\pi }\frac{dz}{z\left( k_{x}\right) },
\end{equation}%
where $z\left( k_{x}\right) =\det \left( \mathcal{B}\left( k_{x}\right)
\right) /\left\vert \det \left( \mathcal{B}\left( k_{x}\right) \right)
\right\vert $. As shown in Figs. \ref{Fig6} (c) and (d) with $h_{z}>\sqrt{%
\left( \mu +2t_{x}+2t_{y}-\mu _{d}\right) ^{2}+\Delta _{s}^{2}}$, the
complex value of $z\left( k_{x}\right) $ varies when $k_{x}$ changes from $0$
to $\pi $, indicating $\left\vert w\right\vert =1$. By considering the
trajectory of $z\left( k_{x}\right) $ in the complex plane as $k_{x}$
changing from $0$ to $\pi $, $z\left( k_{x}\right) $ moves from a point on
the negatively real axis to the positive axis while crossing the imaginary
axis exactly once. It is clear that the winding number $w=-1$ when $%
t_{sox}>0 $ in topological phase, as shown in Fig. \ref{Fig6} (c), and the
winding number $w=+1$ when $t_{sox}<0$ in topological phase, as shown in
Fig. \ref{Fig6} (d). Namely, the sign of winding number for the defect chain
is determined by the sign of SOC in the topological phase.

\section{Low-energy theory of topological superfluids on a defective
rectangle}

\label{App2} Remarkably, from Eq. (\ref{eq2D}) the term $\Delta _{\mathrm{%
edge}}$ doesn't change sign, but the coefficient $\alpha \left( s\right) $
changes sign at two corners of defect-rectangle. This will give rise to a
Majorana pair at the corner where $\alpha \left( s\right) $ changes sign.
Hereafter, we will give the analytic solutions of Majorana corner modes.

According to the particle-hole symmetry of $H_{\mathrm{edge}}$, i.e., $%
\left\{ H_{\mathrm{edge}},\sigma _{y}\tau _{y}\right\} =0$, we have%
\begin{equation}
H_{\mathrm{edge}}\sigma _{y}\tau _{y}=-\sigma _{y}\tau _{y}H_{\mathrm{edge}}.
\label{A1}
\end{equation}%
It can be concluded that if there exist zero-energy states of $H_{\mathrm{%
edge}}$, these states are also eigenstates of $\sigma _{y}\tau _{y}$.
Therefore, we assume the zero-energy wave functions in the \textquotedblleft
edge coordinate" $s$ have the following forms: 
\begin{eqnarray}
\Psi _{0,+} &=&f_{+}\left( s\right) \left\vert y_{+}\right\rangle _{\sigma
}\left\vert y_{+}\right\rangle _{\tau }+g_{+}\left( s\right) \left\vert
y_{-}\right\rangle _{\sigma }\left\vert y_{-}\right\rangle _{\tau }, \\
\Psi _{0,-} &=&f_{-}\left( s\right) \left\vert y_{+}\right\rangle _{\sigma
}\left\vert y_{-}\right\rangle _{\tau }+g_{-}\left( s\right) \left\vert
y_{-}\right\rangle _{\sigma }\left\vert y_{+}\right\rangle _{\tau }.
\end{eqnarray}%
Then we have $\sigma _{y}\tau _{y}\Psi _{0,\pm }=\pm \Psi _{0,\pm }$ and the
Schr\"{o}dinger equation at the corner between the edge \textrm{I} and 
\textrm{II} (corner $s=0$ ) is $H_{\mathrm{edge}}\Psi _{0,+}=0.$ Using the
eigenvector $\Phi =\left( f_{+}\left( s\right) ,g_{+}\left( s\right) \right)
^{T}$, the above equation (\ref{A1}) can be rewritten as%
\begin{equation}
\left( t\partial _{s}^{2}+i\alpha \partial _{s}\sigma _{z}+\mu _{\mathrm{edge%
}}-i\Delta _{\mathrm{edge}}\sigma _{z}+h_{z}\sigma _{x}\right) \Phi =0.
\end{equation}%
We assume $f_{+}\left( s\right) =A_{+}e^{-\eta _{+}s}$, $g_{+}\left(
s\right) =B_{+}e^{-\eta _{+}s}$ ($s>0$) and write 
\begin{equation}
\left( -it\eta _{+}^{2}\sigma _{z}+\alpha \eta _{+}-i\mu _{\mathrm{edge}%
}\sigma _{z}-\Delta _{\mathrm{edge}}+h_{z}\sigma _{y}\right) \tilde{\Phi}=0,
\end{equation}%
with $\tilde{\Phi}=\left( A_{+},B_{+}\right) ^{T}$. According to the
vanishing determinant of the above matrix, we obtain 
\begin{equation}
\eta _{+}=\frac{1}{2}\sqrt{-\frac{2\varkappa }{3}+\delta }-\frac{1}{2}\sqrt{-%
\frac{4\varkappa }{3a}-\delta +\zeta _{+}}>0,
\end{equation}%
where $\varkappa $, $\delta $, $a$, and $\zeta _{+}$ have been written
explicitly in the main text. Then we have $\ A_{+}/B_{+}=e^{i\phi _{+}}$ with%
\begin{equation}
e^{i\phi _{+}}=\frac{h_{z}\left[ -\left( t\eta _{+}^{2}+\mu _{\mathrm{edge}%
}\right) +i\left( \alpha \eta _{+}-\Delta _{\mathrm{edge}}\right) \right] }{%
\left( t\eta _{+}^{2}+\mu _{\mathrm{edge}}\right) ^{2}+\left( \alpha \eta
_{+}-\Delta _{\mathrm{edge}}\right) ^{2}}.
\end{equation}%
At last, we get the MZM $\Psi _{0,+}$ at the corner between the edge \textrm{%
I} and \textrm{II}. Following similar approach as in previous calculations,
we can get another zero energy solution $\Psi _{0,-}=C_{-}e^{-\eta
_{-}\left\vert s\right\vert }\left( e^{i\frac{\phi _{-}}{2}}\left\vert
y_{+}\right\rangle _{\sigma }\otimes \left\vert y_{-}\right\rangle _{\tau
}+e^{-i\frac{\phi _{-}}{2}}\left\vert y_{-}\right\rangle _{\sigma }\otimes
\left\vert y_{+}\right\rangle _{\tau }\right) $ with $C_{-}$ the
normalization constant, where%
\begin{eqnarray}
\eta _{-} &=&\frac{1}{2}\sqrt{-\frac{2\varkappa }{3}+\delta }+\frac{1}{2}%
\sqrt{-\frac{4\varkappa }{3a}-\delta +\zeta _{-}}>0, \\
e^{i\phi _{-}} &=&\frac{h_{z}\left[ -\left( t\eta _{-}^{2}+\mu _{\mathrm{edge%
}}\right) +i\left( \alpha \eta _{-}+\Delta _{\mathrm{edge}}\right) \right] }{%
\left( t\eta _{-}^{2}+\mu _{\mathrm{edge}}\right) ^{2}+\left( \alpha \eta
_{-}+\Delta _{\mathrm{edge}}\right) ^{2}},
\end{eqnarray}%
and the coefficients $\varkappa $, $\delta $, $a$ and $\zeta _{-}$ are
listed in the main text. In summary, there are two Majorana modes (a
Majorana corner pair) localized around the corner with analytic solution
given in Eq. (\ref{eq2DS}). Regarding the corner between \textrm{III} and 
\textrm{IV}, there exists another SOC domain wall and we similarly have a
Majorana pair there.

\section{Low-energy theory of topological superfluids on a defective ring}

\begin{figure}[tbp]
\centering\includegraphics[width=0.48\textwidth]{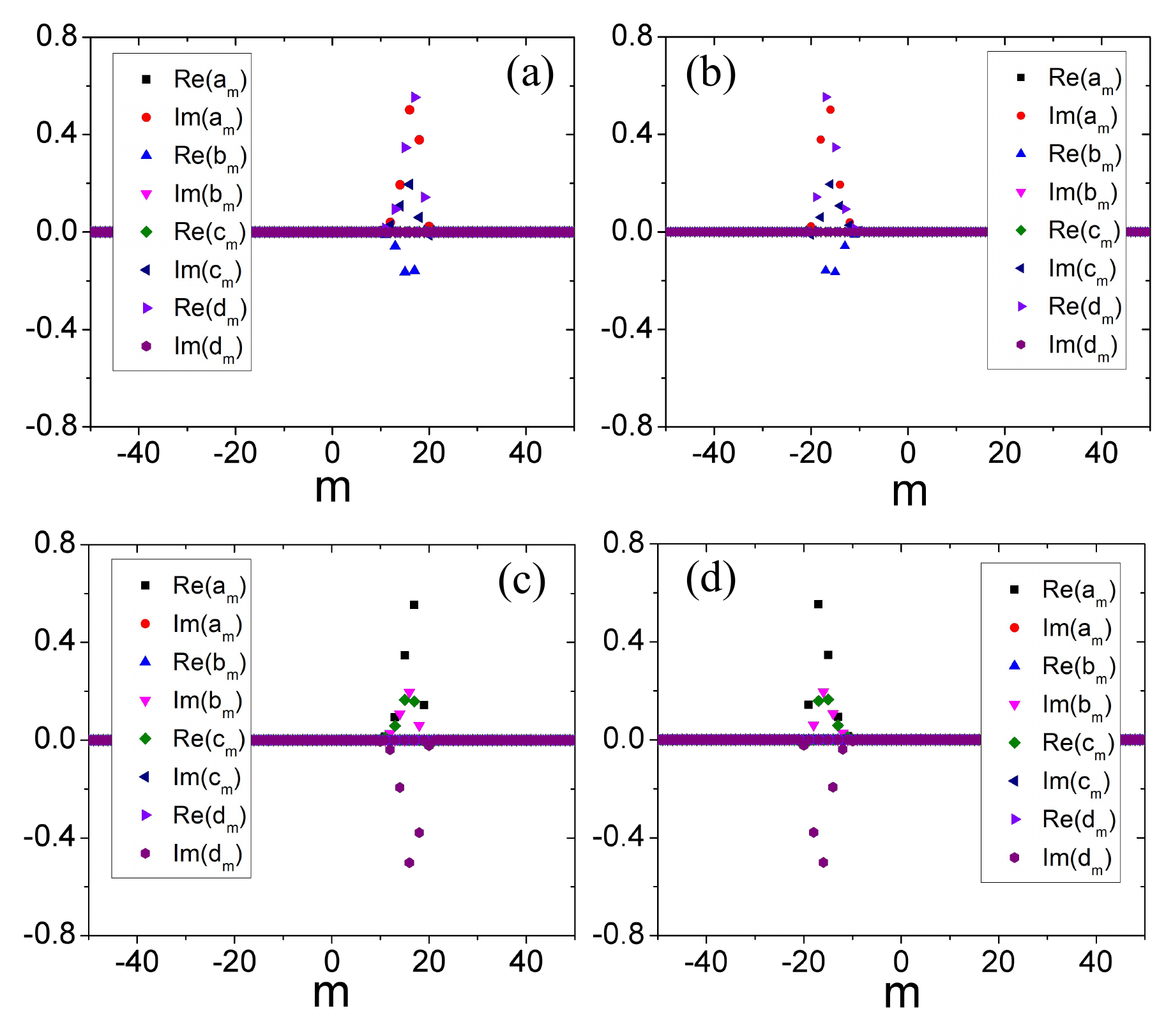}
\caption{The \ real and imaginary parts of coefficients $\left(
a_{m},b_{m},c_{m},d_{m}\right) $ for the wave functions of four Majorana
modes, respectively. Parameters $\tilde{\protect\alpha}_{0}=0.04$, $\protect%
\eta =0.005,\protect\mu ^{\prime }=0.2,h_{z}=-1.4,\Delta =0.8$ are used for
all panels. }
\label{FigS1}
\end{figure}

\label{App3}The general Hamiltonian for a SOC fermi gas with Cooper pairing
is given by%
\begin{equation}
H_{\mathrm{r}}=\frac{1}{2m_{0}}\left( \vec{k}-k_{0}\vec{\sigma}\right)
^{2}-\mu \tau _{z}+h_{z}\sigma _{z}+\Delta _{s}\tau _{x},
\end{equation}%
where $\vec{k}=-i\vec{\nabla}$. Assuming spin $\vec{\sigma}$ is along $x$
and neglecting the constant energy shift $k_{0}^{2}/(2m_{0})$, we get the
Hamiltonian in Eq.~(\ref{eq1a}). In the polar coordinate, 
\begin{equation}
\left( 
\begin{array}{c}
\vec{i} \\ 
\vec{j}%
\end{array}%
\right) =\left( 
\begin{array}{cc}
\cos \phi & -\sin \phi \\ 
\sin \phi & \cos \phi%
\end{array}%
\right) \left( 
\begin{array}{c}
\vec{e}_{\rho } \\ 
\vec{e}_{\varphi }%
\end{array}%
\right) .
\end{equation}%
The Laplace operator in Descartes and polar coordinates is written as%
\begin{equation}
\vec{\bigtriangledown}=\partial _{x}\vec{i}+\partial _{y}\vec{j}=\partial
_{\rho }\vec{e}_{\rho }+\frac{1}{\rho }\partial _{\phi }\vec{e}_{\phi }.
\end{equation}%
Then we have $\vec{k}^{2}=-\vec{\bigtriangledown}\cdot \vec{\bigtriangledown}%
=-\left( \partial _{\rho }^{2}+\frac{1}{\rho ^{2}}\partial _{\phi
}^{2}\right) $ and 
\begin{equation}
\vec{k}\cdot \left( \sigma _{x}\vec{i}\right) =-i\left[ \cos \phi \partial
_{\rho }+\frac{1}{\rho }\partial _{\phi }\left( -\sin \phi \right) \right]
\sigma _{x}.
\end{equation}%
Finally, Eq. (\ref{eq1a}) can be rewritten as the following form:%
\begin{eqnarray}
\mathcal{H} &=&\frac{1}{2m_{0}}\left[ -\left( \partial _{\rho }^{2}+\frac{1}{%
\rho ^{2}}\partial _{\phi }^{2}\right) -ik_{0}\left( \cos \phi \partial
_{\rho }-\frac{1}{\rho }\sin \phi \partial _{\phi }\right) \right.  \notag \\
&&\left. \times \sigma _{x}-ik_{0}\left( \cos \phi \partial _{\rho }-\frac{1%
}{\rho }\sin \phi \partial _{\phi }\right) \sigma _{x}\right] \tau _{z}-\mu
\tau _{z}  \notag \\
&&+h_{z}\sigma _{z}+\Delta _{s}\tau _{x}.
\end{eqnarray}%
Consider ultracold atoms trapped in a ring-shaped trapping potential, where
the radii $\rho $ is fixed. Terms with respect to $\partial _{\rho }$
disappear. After substituting $\eta =\frac{1}{2m_{0}\rho ^{2}},$ $\tilde{%
\alpha}_{0}=\frac{k_{0}}{m_{0}\rho }=\frac{\alpha _{0}}{\rho }$, and $\mu
^{\prime }=\mu -\frac{k_{0}^{2}}{2m_{0}}$, the Hamiltonian $\mathcal{H}$
becomes Eq. (\ref{eq2}).

Because $\mathcal{H}\left( \phi \right) $ is invariant under a $2n\pi $
rotation with $n$ being an integer, we assume the wave functions take
following form as $\Phi \left( \phi \right) =\left( u_{a}\left( \phi \right)
,u_{b}\left( \phi \right) ,u_{c}\left( \phi \right) ,u_{d}\left( \phi
\right) \right) ^{T}$, where $u_{\nu =a,b,c,d}\left( \phi \right)
=\sum_{m=-N}^{N}\nu _{m}e^{im\phi }$. Plugging $\mathcal{H}\left( \phi
\right) $ and $\Phi \left( \phi \right) $ into the Schr\"{o}dinger equation $%
\mathcal{H}\Phi \left( \phi \right) =E\Phi \left( \phi \right) $, and
matching coefficients for $e^{im\phi }$, one can obtain a series of coupled
equations as 
\begin{eqnarray}
Ea_{m} &=&\left( \eta ^{2}m^{2}-\mu ^{\prime }+h_{z}\right) a_{m}+i\frac{%
\tilde{\alpha}_{0}}{2}\left( m-1\right) b_{m-1}  \notag \\
&&-i\frac{\tilde{\alpha}_{0}}{2}\left( m+1\right) b_{m+1}+\Delta _{s}c_{m},
\label{r1} \\
Eb_{m} &=&\frac{i\tilde{\alpha}_{0}}{2}\left( m-1\right) a_{m-1}-i\frac{%
\tilde{\alpha}_{0}}{2}\left[ m+1\right] a_{m+1}  \notag \\
&&+\left( \eta m^{2}-\mu ^{\prime }-h_{z}\right) b_{m}+\Delta _{s}d_{m},
\label{r2} \\
Ec_{m} &=&\Delta _{s}a_{m}+\left[ -\eta m^{2}+\mu ^{\prime }+h_{z}\right]
c_{m}-i\frac{\tilde{\alpha}_{0}}{2}\left( m-1\right) d_{m-1}  \notag \\
&&+i\frac{\tilde{\alpha}_{0}}{2}\left( m+1\right) d_{m+1},  \label{r3} \\
Ed_{m} &=&\Delta _{s}b_{m}-i\frac{\tilde{\alpha}_{0}}{2}\left( m-1\right)
c_{m-1}+i\frac{\tilde{\alpha}_{0}}{2}\left( m+1\right) c_{m+1}  \notag \\
&&+\left[ -\eta m^{2}+\mu ^{\prime }-h_{z}\right] d_{m}.  \label{r4}
\end{eqnarray}%
By solving above coupled equations (\ref{r1})-(\ref{r4}) with the truncation
bounds for $m$ up to $50$, the eigenenergies and corresponding
eigenfunctions $\Phi \left( \phi \right) $ could be obtained. Fig. \ref{Fig5}
(c) in the main text presents the eigenspectrum, indicating that there are
four Majorana zero modes. The coefficients $\nu _{m}$ of wavefunctions of
four Majorana modes are plotted in Fig. \ref{FigS1}.

\end{document}